# A spectrograph instrument concept for the Prime Focus Spectrograph (PFS) on Subaru Telescope

Sébastien Vives*[a], David Le Mignant[a], Fabrice Madec[a], Marc Jaquet[a], Eric Prieto[a], Laurent Martin[a], Olivier Le Fèvre[a], James Gunn[b], Michael Carr[b], Stephen Smee[c], Robert Barkhouser[c], Hajime Sugai[d], Naoyuki Tamura[d]

[a]Laboratoire d'Astrophysique de Marseille (CNRS/Aix-Marseille University), France
[b]Department of Astrophysical Sciences, Princeton University, Princeton, NJ 08544, USA
[c]Department of Physics and Astronomy, Johns Hopkins University, Baltimore, MD, USA
[d]Institute for the Physics and Mathematics of the Universe (IPMU), University of Tokyo (Japan)

## ABSTRACT

We describe the conceptual design of the spectrograph opto-mechanical concept for the SuMIRe Prime Focus Spectrograph (PFS) being developed for the SUBARU telescope. The SuMIRe PFS will consist of four identical spectrographs, each receiving 600 fibers from a 2400 fiber robotic positioner at the prime focus. Each spectrograph will have three channels covering in total, a wavelength range from 380 nm to 1300 nm. The requirements for the instrument are summarized in Section 1. We present the optical design and the optical performance and analysis in Section 2. Section 3 introduces the mechanical design, its requirements and the proposed concepts. Finally, the AIT phases for the Spectrograph System are described in Section 5.

## 1. INTRODUCTION

The Prime Focus Spectrograph (PFS) of the Subaru Measurement of Images and Redshifts (SuMIRe) project [1] has been endorsed by the Japanese community as one of the main future instruments of the Subaru 8.2-meter telescope at Mauna Kea, Hawaii. PFS is a multiplexed fiber-fed optical and near-infrared spectrograph (Nfiber=2400, 380≤ λ ≤ 1300nm), offering unique opportunities in survey astronomy and targeting cosmology with galaxy surveys, Galactic archaeology, and studies of galaxy/AGN evolution [2].

The successful Conceptual Design Review (CoDR), held in March 2012, triggered the decision to commence construction with first light predicted in 2017. The PFS collaboration, led by IPMU in Japan, consists of USP/LNA in Brazil, Caltech/JPL, Princeton, and JHU in USA, LAM in France, ASIAA in Taiwan, and NAOJ/Subaru.

Figure 1 depicts the main components of the Subaru Prime Focus Spectrograph (PFS). It is composed of the Wide Field Corrector (WFC) which corrects the prime focus of the SUBARU telescope for the Prime Focus Instrument (PFI). PFI covers a field-of-view of 1.3deg with 2400 fiber positioners [3]. Then, the Fiber System [4] relays the light captured by each PFI positioner to the Spectrograph System (SpS).

This article gives the current status of the opto-mechanical design of the Spectrograph System (SpS) including developments since the CoDR.

*sebastien.vives@oamp.fr; phone +33 (0)491 056 931

The Spectrograph System (SpS) is composed of four identical modules fed by 600 fibers each. Each module incorporates three channels covering the wavelength ranges 0.38-0.67μm ("Blue"), 0.65-1.00μm ("Red"), and 0.97-1.30μm ("NIR") respectively; with resolving power which progresses fairly smoothly from about 2000 in the blue to about 5000 in the infrared. Table 1 summarizes the main characteristics of the SpS.

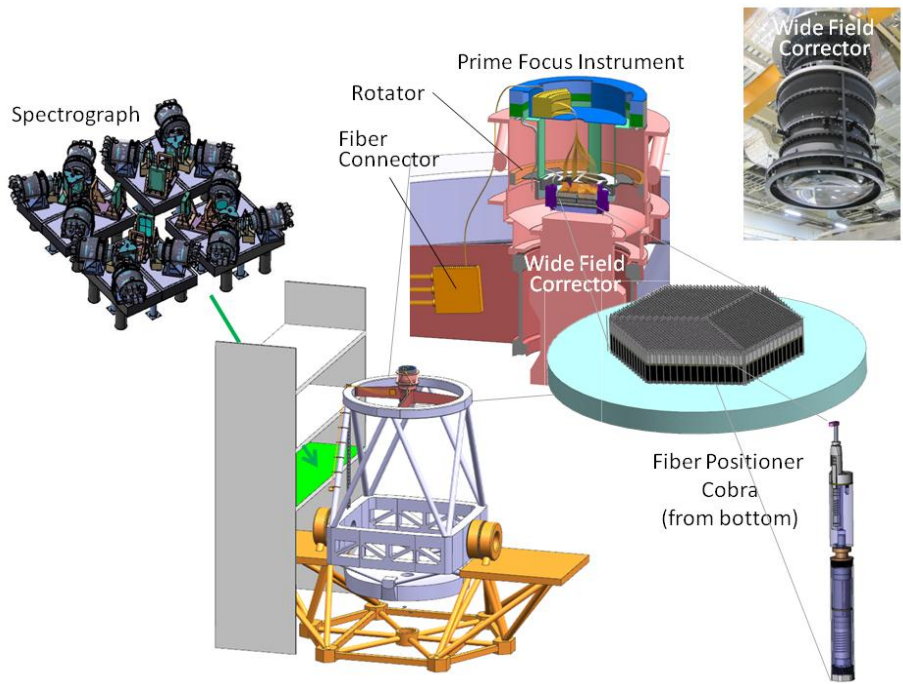

Figure 1. Overview of the Prime Focus Spectrograph (PFS) in the SUBARU telescope.

Table 1. Main characteristics of the PFS instrument.

| **Overview** | |
|---|---|
| Number of fibers | 2400 (600 per module) |
| Location | M3 floor of the SUBARU Telescope |
| **Optics** | |
| Wavelength range and Resolution | Blue: 0.38-0.67μm (R ≥ 2200) <br> Red: 0.65-1.00μm (R ≥ 2900) <br> NIR: 0.97-1.30μm (R ≥ 4200) |
| Fibers core diameter | 128 μm |
| Fiber input F-ratio | F/2.8 (without focal ratio degradation) |
| Collimator focal ratio | F/2.5 |
| Camera focal ratio | F/1.1 |
| VPH Grating dimension | 280mm in diameter |
| Detector format | 4kx4k with 15μm pixel size |
| **Mechanics** | |
| Mass | <10 tons |
| Footprint | 5x5 m |
| **Operating conditions** | |
| Room temperature | 0°C |
| Altitude | 4200m above sea level |

## 2. OPTICAL DESIGN

**2.1 Description**

Figure 2 shows the overall optical layout of one spectrograph module. The proposed optical design is based on a Schmidt collimator facing a Schmidt camera. This architecture is very robust, well known and documented. It allows for high image quality with only few simple elements (high throughput) at the expense of the central obscuration, which leads to larger optics.

The object plane is made of optical fibers arranged along a curved slit, coming out from the telescope primary focus. The beam coming out of the fibers is collimated by a standard Schmidt camera (inverted). Two dichroics are inserted in the beam (to form the three spectral channels, also referred as arms) between the spherical collimating mirror (R=1400mm) and the Schmidt correctors (one per channel). The three Schmidt correctors are identical except for the coating, optimized for each waveband.

Once the linear object is collimated, the beam is dispersed by a VPHG assembly, one per arm, in the perpendicular direction with respect to the object direction. VPH gratings offer high efficiency and their use in transmission allows the camera to be placed close to the pupil, which minimized the size of the camera optics. Nominal line densities for the blue, red, and infrared are 676 lines/mm, 554 lines/mm, and 571 lines/mm, respectively. Preliminary theoretical performance prediction based on Rigorous Couple Wave Analysis (RCWA) has been provided by Kaiser Optical Systems, Inc. (KOSI) for each grating. All three designs produce theoretical efficiencies in excess of 90%.

For each of the three channels, a camera images the dispersed line on a square detector (~60x60mm). For the blue and red channels, the detector will be a pair of close butted 2Kx4K, 15 micron pixel CCDs to yield a 4Kx4K format, while for the IR channel, a 4Kx4K 15 micron pixel; 1.75 micron cutoff Mercury-Cadmium-Telluride device will be used. See [5] for more details on the detectors.

Each camera is based on a Schmidt concept adapted to the fast focal ratio (F/1.1) and to the large field of view: in particular, the classical spherical mirror has been replaced by a Mangin-like mirror (i.e. a meniscus lens with the reflective surface on the rear side of the glass). Note that, strictly speaking, it is not a "Mangin mirror" since the meniscus is not negative in our case. As shown Figure 3, on top of the Mangin-like mirror, each camera is composed of a Schmidt corrector, and a field corrector (to flatten the nominal curved image plane provided by the Schmidt system). The Schmidt corrector is made of Silica and has its two surfaces aspherical. The Mangin-like mirror is composed of a spherical meniscus in Silica of 40 mm thickness at the vertex with the reflective coating deposited on the rear surface. Therefore the light passes twice in the glass of the mirror. The radii of curvature of the two surfaces are very close: 690mm and 670mm for the front and the back surfaces, respectively.

The proposed optical design has been optimized to achieve the requested image quality while simplifying the manufacturing of the whole optical system. In particular:

- The number of identical elements in the three channels has been maximized: all correctors of the collimator and all Mangin-like mirrors are identical in the three channels;
- The number of aspherical components has been minimized (4 over 9 optical surfaces).

This architecture (i.e. a Schmidt collimator facing a Schmidt camera) is very robust, well known and documented. It allows for high image quality with only few simple elements (high throughput) at the expense of the central obscuration, which leads to larger optics. One of the main driver for the design was to maintain the size of the volume-phase holographic grating (VPHG) in the current standard manufactory capability <300mm in diameter.

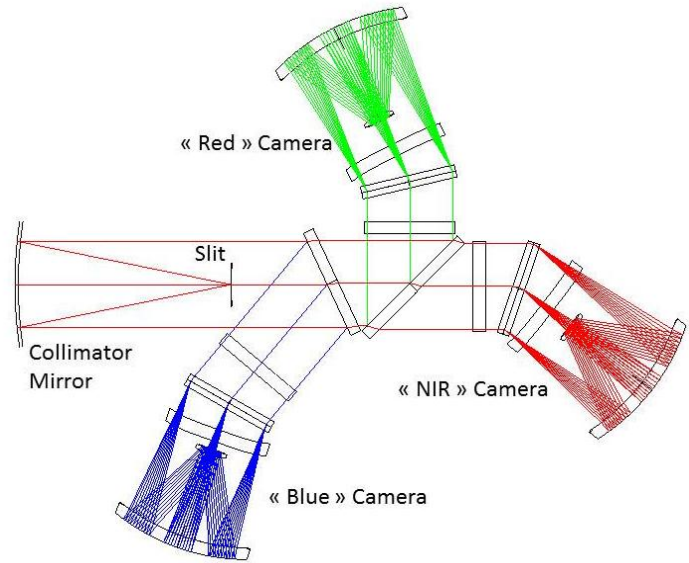

Figure 2. Optical Layout of one module of the Spectrograph System. Each module is identical.

Regarding the manufacturing aspects, the Schmidt correctors of the cameras appear to be the most (and unique) critical components in this design. With two surfaces aspherical, the sag departure from the best sphere can reach 2mm making the inspection of such highly aspheres really challenging. It is possible to reduce the deformation (below 0.5mm) by splitting the corrector in two lenses (with only one aspherical surface per lens). This option will be investigated further.

Because of the F/1.1 beam, the depth in focus at the focal plane is very short and the location (along the optical axis) of the detector is critical. In order to relax the positioning of the detector, a fine adjustment mechanism will be implemented.

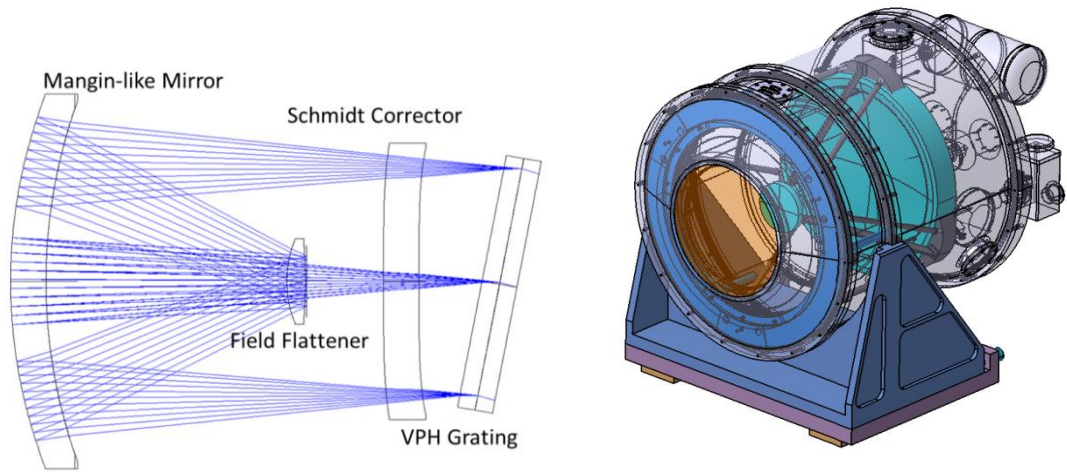

Figure 3. Optical Layout (left) and mechanical design (right) of the camera optics of the SpS.

### 2.2 Performance

The proposed optical design is compliant with the specifications of image quality expressed in ensquared energy within 3 and 5 pixels. Figure 4 represents the spot matrix (field versus wavelength) in the three channels.

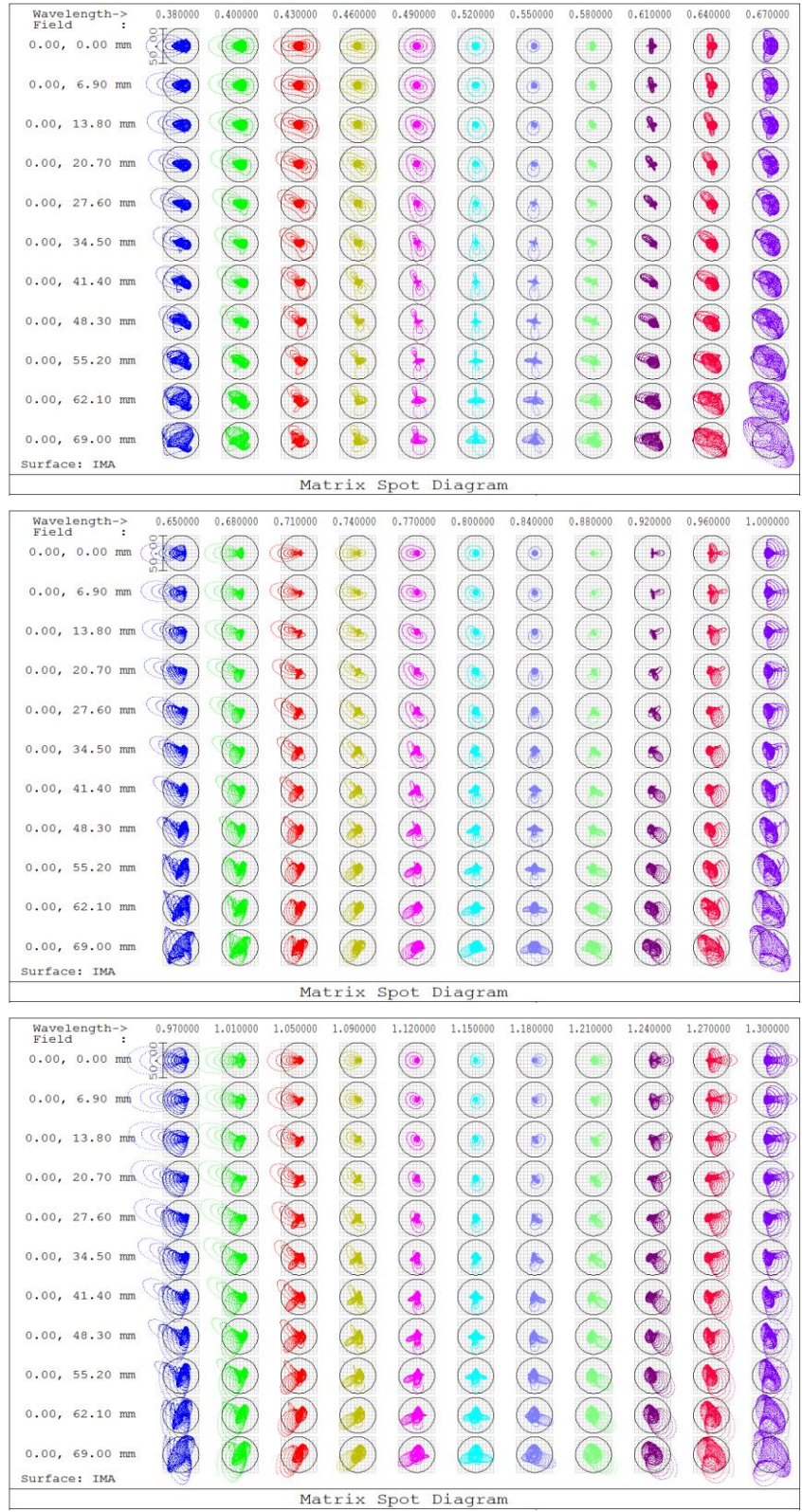

Figure 4. Spot Matrix (matrix field versus wavelength) in the "Blue" (top), "Red" (middle), and "NIR" (bottom). The circle corresponds to the dimension of the fiber on the detector (i.e. 50µm).

# 3. MECHANICAL DESIGN

## 3.1 Location

The primary candidate for the location of the SpS is on the third floor at the "IR" side in the dome. This floor has not been used or even prepared for any scientific instruments. Hence, the installation of the SpS will require major upgrades. In particular, the floor will have to be fully modified so that it can support the instrument and necessary equipment for instrument commissioning and operations.

The size of the biggest clearance existing on the floor is 5.6 meters by 6.3 meters. Figure 1 also shows a possible implementation of four spectrograph modules with a total footprint of about 5x5m. The clear space between each bench is about 0.5m.

## 3.2 Description

The SpS has to be modular in its design to allow for Assembly, Integration and Tests (AIT) and for its safe transport up to the summit. This is the main driver for the mechanical design. In particular, the SpS will be firstly fully integrated and validated at LAM (France) before it is shipped to Hawaii. All sub-assemblies will be indexed on the bench to allow for their accurate repositioning.

Figure 5 shows the different components of one Spectrograph Module. They are mounted on a 2x2m optical bench. This bench allows us to cope with both the transport constrains (truck and container dimensions); and the limitations of the access to the SUBARU telescope M3 floor. All elements (except for the cryostats) will be baffled to prevent for the external stray light and dust to enter.

For integration and maintenance reasons, the fiber slit, the collimator mirror (with the Schmidt correctors) and the dichroics will be mounted on a common bench (not represented in Figure 5).

The cryostat assembly will provide the required vacuum and thermal environment for the detector and the camera optics. The Schmidt corrector of the camera will act as the entrance window of the vessel. A hexapod mount will maintain the camera optics in the vessel. The hexapod struts will be made of Invar + G10 (with flexure in TA6V) in order to limit the distance variation between the Schmidt corrector and the Mangin-like mirror under cryogenic conditions, while minimizing the thermal exchange with the struts.

A spider attached to the Mangin-like mirror supports the detector and field flattener using Invar inserts. The spider struts are made of two superimposed materials minimizing the vignetting: Invar to limit the focus variation (between ambient and cryogenic conditions), and Copper to cool-down the detector (thermal conductivity).

The detector is also mounted on a tip/tilt and focus adjustment mechanism of limited range to accomplish final alignment and focus. The adjustments are made via flexures with long, thin lever arms to a stepping-motor/screw mechanism near the outer wall of the vessel.

More details on the cryostat, the detector and the focus mechanism are given in [5].

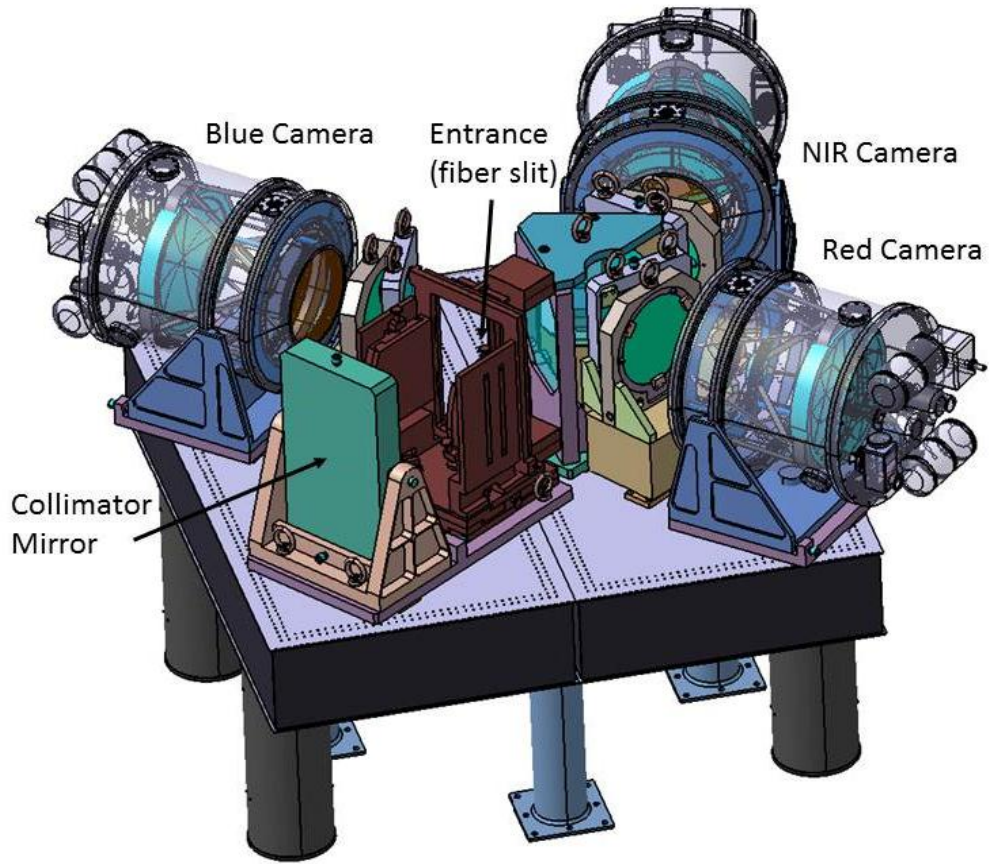

Figure 5. Spectrograph Module mechanical overview. The baffling covering the bench is not drawn.

### 3.3 Assembly Integration and Test (AIT)

The four modules of the SpS will be firstly integrated and fully validated at LAM. All sub-assemblies have to be validated before their final integration on a module. Indeed, the sub-assemblies will be developed in parallel (in France but also in Brazil and in the USA), while the modules will be integrated in a sequence.

We also plan to use dedicated AIT stations since most of the operations will have to be repeated several times. This approach allows guaranteeing a good homogeneity in the AIT tests and validation (i.e. from one spectral channel to another), but also allows securing the handling of heavy and/or fragile components.

We plan to minimize the AIT activities at the telescope, as it is difficult to work at this high altitude.

We currently study options for a Spectrograph development where an industrial partner will further study our design, manufacture, test and validate the individual components for each of the 12 arms, then further integrate them in units that then be assembled at LAM.

## 4. CONCLUSION

We have presented the Conceptual Design study for the PFS-SUMIRe spectrographs. This study allows demonstrating the development feasibility of the Spectrograph System and we believe we have now identified a baseline configuration. We will use this baseline to further work on the spectrograph development plan with an industrial partner and our academic partners later in 2012/2013. The first spectrograph arm should be aligned in the LAM by early 2014.

## ACKNOWLEDGMENTS

We gratefully acknowledge support from the Funding Program for World-Leading Innovative R&D on Science and Technology (FIRST) "Subaru Measurements of Images and Redshifts (SuMIRe)", CSTP, Japan.